\PassOptionsToPackage{table}{xcolor}
\documentclass{lightspeed}

\definecolor{titlebg}{HTML}{EEF3FA}

\renewcommand{\title}[1]{\newcommand{\titlelist}{{\LARGE\sffamily\bfseries #1}}}
\renewcommand\authorformat[2][]{{\sffamily\bfseries\color{metabg}#2$^{#1}$}}
\renewcommand\affiliationformat[2][]{{\small\color{metabg}$^{#1}$#2}}
\renewcommand\contributionformat[2][]{{\small\color{metabg}$^{#1}$#2}}
\renewcommand\emailformat[2][]{{\footnotesize\texttt{\href{mailto:#2}{#2}}}}
\renewcommand\metadataformat[2][]{{\small{\sffamily\bfseries #1:} #2}}

\renewcommand{\mymaketitle}{%
  \begingroup
  \setlength{\parindent}{0pt}
  \setlength{\parskip}{0pt}
  \vspace*{-0.35cm}
  \begin{tcolorbox}[enhanced,colback=titlebg,colframe=titlebg,
      boxrule=0pt,arc=9pt,left=0.55cm,right=0.55cm,
      top=0.45cm,bottom=0.40cm,before skip=0pt,after skip=0.45cm]
    {\raggedright\nohyphens
      {\setstretch{1.08}\titlelist\par}
      \vspace{0.28cm}
      {\authorlist\par}
      \vspace{0.14cm}
      {\affiliationlist\par}
      \ifdefempty{\contributionlist}{}{\vspace{0.06cm}\contributionlist\par}
      \ifdefempty{\emaillist}{}{\vspace{0.10cm}{\color{metabg}\emaillist\par}}
      \vspace{0.32cm}
      {\small\setstretch{1.08}\color{metafg}\abstractlist\par}
      \ifdefempty{\headercontentlist}{}{\vspace{0.26cm}\headercontentlist}
      \ifdefempty{\metadatalist}{}{\metadatalist\par}
    }
  \end{tcolorbox}
  \endgroup
  \FloatBarrier
}

\microtypesetup{expansion=false}
\setcitestyle{square,comma,numbers,sort&compress}

\usepackage[T1]{fontenc}
\usepackage[utf8]{inputenc}
\usepackage{microtype}
\usepackage{inconsolata}

\usepackage{graphicx}
\usepackage{wrapfig}
\usepackage{needspace}
\usepackage{placeins}
\usepackage{flafter}
\usepackage{booktabs}
\usepackage{multirow}
\usepackage{amsmath}
\usepackage{amssymb}
\usepackage{url}
\usepackage{enumitem}
\usepackage{xspace}

\newcommand{\sysname}{\textsc{DeskIF-Bench}\xspace}
\newcommand{\methodname}{\textsc{ANIS}\xspace}

\definecolor{decS}{HTML}{A6D5CB}
\definecolor{decM}{HTML}{C6E3DC}
\definecolor{decL}{HTML}{E2F0EC}
\definecolor{incL}{HTML}{F7E0CE}
\definecolor{incM}{HTML}{EDBB94}
\newcommand{\code}[1]{\texttt{#1}}
\newcommand{\brk}{\allowbreak}
\colorlet{takeawaybg}{titlebg}
\newtcolorbox{takeawaybox}{
  enhanced,
  breakable,
  colback=takeawaybg,
  colframe=takeawaybg,
  boxrule=0pt,
  arc=0.8mm,
  left=1.2mm,
  right=1.2mm,
  top=0.9mm,
  bottom=0.9mm
}
\newcommand{\takeaway}[1]{\begin{takeawaybox}\textbf{Takeaway.}\ #1\end{takeawaybox}}

\newcommand{\vch}[3]{1.00$\rightarrow$#1$\rightarrow$#2$\rightarrow$#3}
\newcommand{\mDn}{\,$\blacktriangledown$}
\newcommand{\mUp}{\,$\blacktriangle$}
\newcommand{\mNo}{\,\phantom{$\blacktriangle$}}

\newcommand{\cNo}[3]{\scriptsize\vch{#1}{#2}{#3}\mNo}
\newcommand{\cDl}[3]{\cellcolor{decL}\scriptsize\vch{#1}{#2}{#3}\mDn}
\newcommand{\cDm}[3]{\cellcolor{decM}\scriptsize\textbf{\vch{#1}{#2}{#3}}\mDn}
\newcommand{\cDs}[3]{\cellcolor{decS}\scriptsize\textbf{\vch{#1}{#2}{#3}}\mDn}
\newcommand{\cUl}[3]{\cellcolor{incL}\scriptsize\vch{#1}{#2}{#3}\mUp}
\newcommand{\cUm}[3]{\cellcolor{incM}\scriptsize\textbf{\vch{#1}{#2}{#3}}\mUp}

\title{A Decade in a Day: Synthesizing Agent-Native Interfaces\\from Human-Oriented Software Evolution}

\author[1,*]{Alan Wu}
\author[2,\dagger]{Wei Tao}
\author[1,\dagger]{Yiwen Guo}

\affiliation[1]{Independent Researcher}
\affiliation[2]{LIGHTSPEED}
\renewcommand{\affiliationlist}{%
  \affiliationformat[1]{Independent Researcher}\quad
  \affiliationformat[2]{LIGHTSPEED}%
}

\email{799167097@qq.com}
\email{wtao@ieee.org}
\email{guoyiwen89@gmail.com}

\abstract{Large language model agents are emerging as an important new class of users of desktop applications. However, existing desktop interfaces were designed for human end users or application developers, leaving agents to interpret visual controls or discover and compose low-level software operations. This mismatch can hinder even capable agents, yet it remains unclear what interface properties make desktop software easier for agents to use. To study this question, we treat software evolution as a natural source of interface variation and introduce \sysname, a benchmark of 255 artifact-verified tasks across six desktop applications. By holding the task, agent, and verifier fixed while varying the software release, we measure how interface changes affect agent performance. Across two models and six applications, the newest release consistently outperforms the oldest. Trajectory analysis identifies three relevant interface properties: discoverability, granularity, and observability. Guided by these properties, we develop \methodname, a framework that synthesizes agent-native interfaces from the capabilities of installed applications. On an expanded evaluation of 368 tasks across eight applications, 
atomic interfaces improve success over native SDK use by 3.8 and 3.5 percentage points with DeepSeek V4 Flash and Claude Haiku 4.5, respectively, while reducing online token usage by approximately 73\%. Goal-level commands further raise success to 74.5\% and 67.4\%, respectively.
}
\headercontent{%
  {\small{\sffamily\bfseries Code:} \texttt{\href{https://github.com/co-evolve-lab/ANIS}{github.com/co-evolve-lab/ANIS}}\par}
  {\small{\sffamily\bfseries Data:} \texttt{\href{https://huggingface.co/datasets/SimAlan/DeskIF-Bench}{huggingface.co/datasets/SimAlan/DeskIF-Bench}}\par}%
}
\date{October 2026}

\begin{document}
\thispagestyle{firstheader}
\maketitle
\begingroup
\renewcommand{\thefootnote}{\fnsymbol{footnote}}
\footnotetext[1]{Work done during an internship at LIGHTSPEED.}
\footnotetext[2]{Corresponding authors.}
\endgroup
\setcounter{footnote}{0}

\section{Introduction}

Large language model (LLM) agents are emerging as a new class of users of desktop applications, performing tasks from document editing to image manipulation~\citep{bonatti2025windows,DBLP:journals/corr/abs-2503-15661,DBLP:journals/corr/abs-2502-08047,li2026advances}. This creates a new interface-design challenge for desktop applications.

Desktop applications expose functionality through interfaces built for human end users and developers, not for LLM agents. Graphical user interfaces (GUIs) assume that users can visually locate controls and interpret continuous visual feedback. Software development kits (SDKs) assume that developers can search documentation, understand an object model, write code, and debug low-level calls \citep{lu2025axis,song2025beyondbrowsing}. 
Agents therefore need to perform the work these interfaces expect of humans: repeatedly interpreting and grounding GUI controls or discovering and composing low-level SDK operations.
Command-line interfaces (CLIs) offer a higher-level alternative through concise, structured, task-oriented commands \citep{demasi2026terminal,yang2026clianything}, but most desktop applications expose no general-purpose CLI, and available CLIs often cover only a narrow subset of application functionality, such as import and export \citep{wang2026declarative,zhou2026guivscli}.

Together, these limitations leave agents without a broadly applicable interface aligned with their capabilities. An interface mediates every action an agent takes on an application. If it neither exposes the capabilities required for a task nor allows them to be composed from available operations, even a capable agent cannot complete the task. A poorly aligned interface therefore becomes the Achilles' heel of agents operating desktop software. This raises a fundamental question: What constitutes a good interface for agents interacting with desktop applications?

Answering this question requires interface variations that are both realistic and controlled, yet hand-crafted variants would merely encode our own assumptions about what agents need. Our key observation is that software evolution has already produced such variations: across releases, an application's core functionality remains largely stable while its interface is continually reshaped by developers responding to real usability pressure. Version history thus offers a ready-made natural experiment: by fixing the agent and tasks and varying only the historical interface, we can attribute performance shifts to documented interface changes.

\begin{figure}[t]
  \centering
  \includegraphics[width=\linewidth]{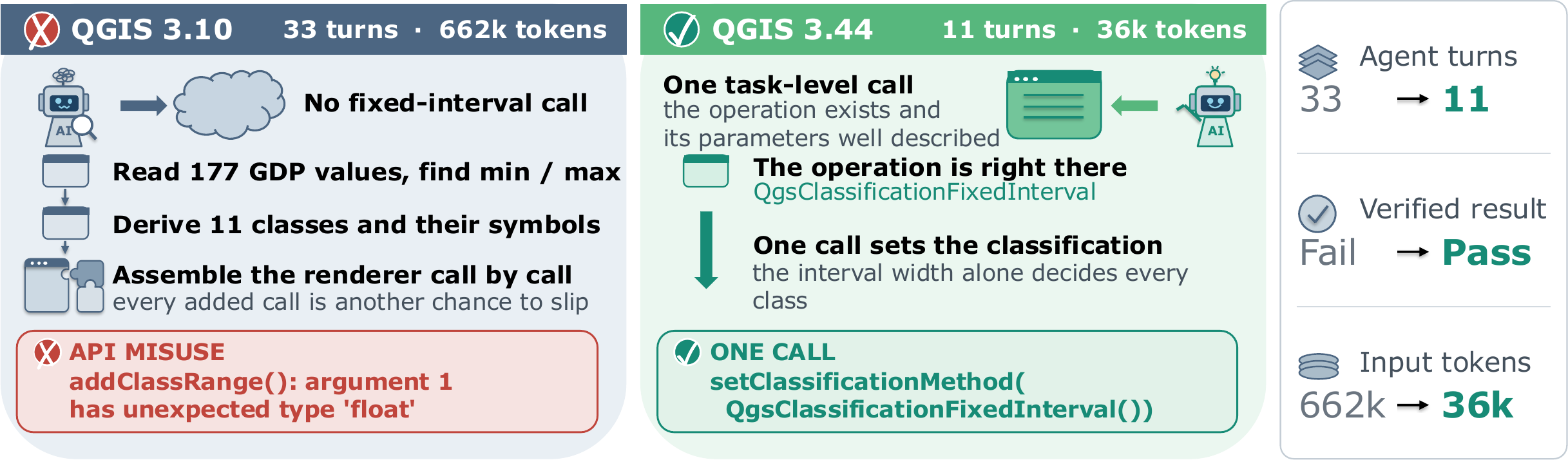}
  \caption{A clearly described fixed-interval operation in QGIS replaces eleven manually constructed ranges, turning failure into success.}
  \label{fig:qgis-case}
\end{figure}

As a concrete illustration, Figure~\ref{fig:qgis-case} compares two historical releases of QGIS, a geographic information system~\citep{qgis2026}, on the same task: creating a graduated map of country GDP with fixed-width intervals. In QGIS 3.10, the agent must derive eleven classes and assemble the renderer through a long chain of poorly documented low-level calls. The long chain invites parameter errors that the sparse SDK descriptions make hard to avoid or repair, and the agent fails after 33 turns and 662k input tokens. QGIS 3.44 instead offers clear SDK descriptions and a task-level \code{QgsClassification\brk FixedInterval} operation that generates the complete classification from a single interval width. The same agent succeeds in 11 turns with 36k input tokens. Clearer operation descriptions and a task-level abstraction make the interface more agent-friendly.

To systematically understand what makes interfaces more agent-friendly, we build \sysname (\emph{Desktop Interface Benchmark}), a benchmark comprising 255 artifact-verified tasks across six applications and five domains. Holding the agent scaffold, backbone model, task setup, and artifact verifier fixed, we vary only the interface exposed to the agent across software releases. The results reveal that the newest release outperforms the oldest in all 12 model--application series, but by uneven margins. From QGIS 3.10 to 3.44, the success rate rises from 33\% to 65\% while mean SDK operations per task fall from 30.4 to 21.2. Impress gains 19 points, from 33\% to 52\%.

Across applications, versions, and trajectories, our analysis identifies three recurring interface properties that help explain these performance differences (Section~\ref{sec:study-explanations}):
\begin{itemize}[leftmargin=*,itemindent=0pt,labelsep=0.35em,topsep=1pt,itemsep=0pt,parsep=0pt]
\itemsep0em
  \item \textbf{Discoverability}: The interface makes supported operations discoverable at runtime and describes their semantics and arguments clearly enough for agents to understand how to invoke them. Poor discoverability forces agents to search documentation and probe live objects before they can act.
  \item \textbf{Granularity}: The interface provides operations aligned with recurring user goals, reducing low-level calls. For example, QGIS 3.44 introduces task-level operations that reduce such overhead.
  \item \textbf{Observability}: The interface exposes the resulting artifact state at the same level at which it is modified, so agents can verify whether the requested outcome has been achieved. Such visibility is particularly important in text-based interaction, where agents lack the visual feedback.
\end{itemize}

Guided by these properties, we design \methodname, an Agent-Native Interface Synthesizer (Section~\ref{sec:method}). \methodname synthesizes a base interface once: application-specific collectors recover candidate operations and properties from complementary sources in the running application, retaining only capabilities whose persistent effects can be verified. Each admitted capability becomes an atomic command with a uniform specification, executable implementation, and observable effect. For discoverability, \methodname groups commands by artifact objects and exposes concise summaries before full specifications. For verification, it pairs writable properties with read commands. 
Because useful higher-level operations depend on goals emerging after deployment, \methodname extends the interface on demand by composing atomic commands into goal-level commands, released only when replay on a fresh artifact reproduces the expected saved state.

We evaluate \methodname on an expanded \sysname benchmark. To test whether the method generalizes beyond the six applications in our controlled cross-version study, we add two applications and collect their tasks using the same pipeline. On this expanded benchmark, the atomic-command interface achieves a success rate of 72.3\%, compared with 68.5\% for direct SDK use and 17.4\% for CLI-Anything. Adding goal-level commands further raises success to 74.5\%. 

In summary, we make the following three contributions:
\begin{itemize}[leftmargin=*]
\itemsep0em
  \item We introduce \sysname, a 255-task benchmark across six applications for controlled cross-version analysis, expanded to 368 tasks across eight applications to assess \methodname's generalization.
  \item We identify three properties from human design knowledge embedded in software evolution: discoverability, granularity, and observability. We derive them by tracing how interface changes across historical releases affect agent task completion.
  \item We design \methodname, an Agent-Native Interface Synthesizer that operationalizes these properties through runtime-grounded atomic commands, verified goal-level compositions, and artifact readback. On the expanded benchmark, it outperforms direct native SDK use and CLI-Anything by 6.0 and 57.1 percentage points, respectively.
\end{itemize}

\section{Related Work}
\label{sec:related-work}

\paragraph{Computer-use agents and benchmarks.}
OSWorld and Windows Agent Arena evaluate agents on realistic desktop tasks with execution-based checkers \citep{xie2024osworld,bonatti2025windows}, while Agent S studies agents that operate the same screen as a person \citep{agashe2025agents}. Recent systems improve this interaction with reusable shortcuts, memory, or hybrid GUI--API control \citep{wang2025mobileagente,jiang2025appagentx,zhang2025ufo2,zhao2025masbench}. These works typically hold the interface fixed and vary the agent. We instead hold the task, agent, and verifier fixed and vary the interface, isolating interface design as the experimental variable.

\paragraph{Interfaces for agents.}
Recent work combines GUI and API control or replaces low-level actions with declarative abstractions \citep{zhang2025apigui,song2025beyondbrowsing,lu2025axis,wang2026declarative}. Work around the Model Context Protocol (MCP) mainly makes existing operations easier for agents to discover and invoke through better schemas, validation, selection, and loading \citep{ni2025doc2agent,agarwal_automated_2025,bandlamudi_framework_2025,sigdel_schema_2026,sadani2026toolattention}. We instead synthesize verified atomic operations, expose separate commands for reading artifact state, and return actionable errors. We also test whether frozen operations can form validated goal-level sequences absent from the native interface. The released atomic result is transport-agnostic and can be exposed via CLI or MCP. Publishing the goal-level sequences remains future work.

\section{Benchmark Construction}
\label{sec:dataset}

We build \sysname{} to study how agent performance changes across desktop-application interfaces exposed by different software releases. 
Specifically, its design follows three criteria:

\begin{itemize}[leftmargin=*,itemindent=0pt,labelsep=0.35em,topsep=1pt,itemsep=0pt,parsep=0pt]
\itemsep0em
  \item \textbf{Execution-based verification.} Each task carries a script verifier, which inspects the artifact the agent saved and returns pass or fail through deterministic checks.
  \item \textbf{Interface-independent task outcomes.} Each task specifies an application-level artifact outcome rather than a particular transient interface state, allowing reuse across releases and interfaces.
  \item \textbf{Diverse applications.} \sysname{} covers six applications across five domains, and every application carries at least 30 tasks, providing substantial coverage for every selected application.
\end{itemize}

\subsection{Software Selection}
\label{sec:dataset-software}

\sysname{} covers desktop applications across multiple domains rather than a single application family. We target five domains: office productivity, 2D graphics editing, 3D content creation, geospatial analysis, and computer-aided design, which span diverse artifact representations, including symbolic documents and slide decks, scalable vector graphics, 3D scenes, layered geospatial data, and parametric CAD models. Within each domain we follow prior benchmarks \citep{xie2024osworld,opencomputer2026} and select widely used applications with long release histories, providing both practical relevance and substantial interface evolution. The result is six applications: Writer and Impress for office productivity, Inkscape for 2D graphics editing, Blender for 3D content creation, QGIS for geospatial analysis, and FreeCAD for computer-aided design.
Version selection criteria and the evaluated releases are provided in Appendix~\ref{sec:appendix-software-versions}.

\subsection{Task Construction}
\label{sec:dataset-tasks}

Tasks in \sysname{} come from two sources: tasks reused from previous datasets, and tasks we synthesize. For applications already covered by previous datasets \citep{xie2024osworld, opencomputer2026}, we filter their tasks and retain those satisfying our criteria (Appendix~\ref{sec:dataset-reuse}). 
Existing benchmarks provide uneven coverage across applications and task types. After filtering, several applications have fewer than 20 reusable tasks, and some applications have none.
We therefore synthesize additional tasks with a pipeline.

Each synthesized task needs an instruction, a reference artifact, and an executable verifier, all produced by an \emph{authoring agent}. The \emph{solving agent} is the agent under evaluation: it receives only the instruction and must produce an artifact satisfying the instruction through the application's interface.
The authoring agent must not only pose a task but also construct a valid solution, since the reference artifact and the verifier depend on that solution. Its capabilities therefore constrain the difficulty of the tasks it can reliably construct. Without additional capability, synthesis would tend to remain within the same capability envelope as the agents under evaluation, limiting benchmark headroom and discriminative power. Following prior work \citep{zheng2026ebench}, we introduce this capability \emph{asymmetry} in two ways:
\begin{itemize}[leftmargin=*]
\itemsep0em
  \item \textbf{Skills}: Where available, the authoring agent receives application-specific skills that provide procedural knowledge of the target software, whereas the solving agent receives no such application-specific skills and must work through the exposed interface.
  \item \textbf{Libraries}: The authoring agent may use a privileged route (e.g., trimesh for Blender) to build the reference artifact and verifier, whereas the solving agent must use the target application.
\end{itemize}

\paragraph{Task proposal.}
Following prior works \citep{xie2024osworld,opencomputer2026}, we ground task proposals in real-world user needs and workflows drawn from official tutorials and community discussions. The authoring agent converts each source into an application-level user goal and formulates an instruction specifying the desired outcome without prescribing interface-specific interactions. 
We retain only proposals that specify artifact-level outcomes rather than transient screen states, and deduplicate them against both reused and synthesized tasks. A proposal is accepted only after the authoring agent constructs a reference artifact and executable verifier.
We additionally require each task outcome to be feasible in every evaluated release. 
We exclude a proposal only when the required artifact cannot be produced through a privileged reference procedure in a particular release, indicating a release-level functional defect rather than an interface limitation.
This feasibility screening reduces confounding from functionality unavailable independently of the agent-facing interface.

\paragraph{Task verification.}
Every accepted task has a dedicated verifier that checks whether the final artifact satisfies the instruction's task-specific requirements.
We use executable checks rather than LLM-as-judge, whose verdicts can vary with the prompt and observation context and are therefore difficult to reproduce consistently \citep{opencomputer2026}. Our verifiers instead use fixed, deterministic checks, including structured artifact inspection and, where the target is inherently visual, pixel-level comparison under fixed tolerances. The same verifier and thresholds are applied unchanged across all evaluated releases of a task. We validate each verifier by confirming that it accepts the reference artifact and rejects task-relevant perturbations that violate the required output conditions. Because these checks operate on explicit artifact properties, their decisions can also be manually audited.

\paragraph{Benchmark summary.}
Table~\ref{tab:benchmark-stats} summarizes the resulting benchmark. Its 255 tasks cover six applications across five domains, and every application carries at least 30 tasks. Of these, 102 are retained from OSWorld and OpenComputer after filtering, and 153 are synthesized to expand coverage in applications and task types insufficiently represented by the reused sets.

\section{Empirical Study}
\label{sec:study}

Because every task in \sysname specifies an artifact-level outcome, ships with its own executable verifier, and is screened for feasibility across evaluated releases, the same task can be posed unchanged against different releases of the same application. We use this property to run an empirical study of interface evolution. Section~\ref{sec:study-setup} states the setup, Section~\ref{sec:study-versions} measures how agent success changes when only the release changes, and Section~\ref{sec:study-explanations} traces those changes to concrete interface properties and, from the same trajectory audit, reports one further property that remains limited across releases. Section~\ref{sec:method} turns these properties into requirements for a synthesized interface.

\subsection{Experimental Setup}
\label{sec:study-setup}

\paragraph{Agent scaffold.}
Because agent performance can depend substantially on the scaffold~\citep{vats2026scaffold,rombaut2026scaffold}, we use a single scaffold that is both open-source and model-agnostic, allowing it to be pinned to an auditable version and used unchanged across backbones. OpenCode~\citep{opencode2026} satisfies these criteria and is used for all runs.

\paragraph{Backbone models.}
To avoid relying on a single model family, we evaluate one proprietary and one open-weight backbone: Claude Haiku 4.5~\citep{anthropic2025haiku45} and DeepSeek V4 Flash~\citep{deepseek2026v4}. Their overall success rates on \sysname are also comparable, reducing confounding from large capability differences. Both models are evaluated across four releases of six applications. Appendices~\ref{sec:appendix-model-performance} and~\ref{sec:appendix-token-distribution} report additional model and documentation analyses from prior experiments.

\paragraph{Cross-version protocol.}
For each backbone model, we vary only the installed application release while holding the task, seed files, agent scaffold, tool permissions, time budget, and verifier fixed across releases. The agent accesses the application only through the programmatic interface exposed by that release, isolating cross-version differences from the agent harness and evaluation procedure.

\paragraph{Evaluation metrics.}
We report success rates using pass@3 to reduce sensitivity to stochastic variation across agent runs: each task is attempted independently three times from the same initial state and counts as successful if at least one attempt passes its artifact verifier, and overall rates aggregate applications by task count. For the trajectory-level analyses in Table~\ref{tab:version-interface}, we restrict the comparison to model--task pairs with verified success in all four releases. Failed trajectories can terminate at arbitrary stages, making their exploration turns and chain lengths incomparable to those of completed solutions. We therefore report these metrics only within this fixed-success cohort.

\subsection{Performance Across Software Versions}

\label{sec:study-versions}

\begin{figure}[!htbp]
\centering
\includegraphics[width=\linewidth]{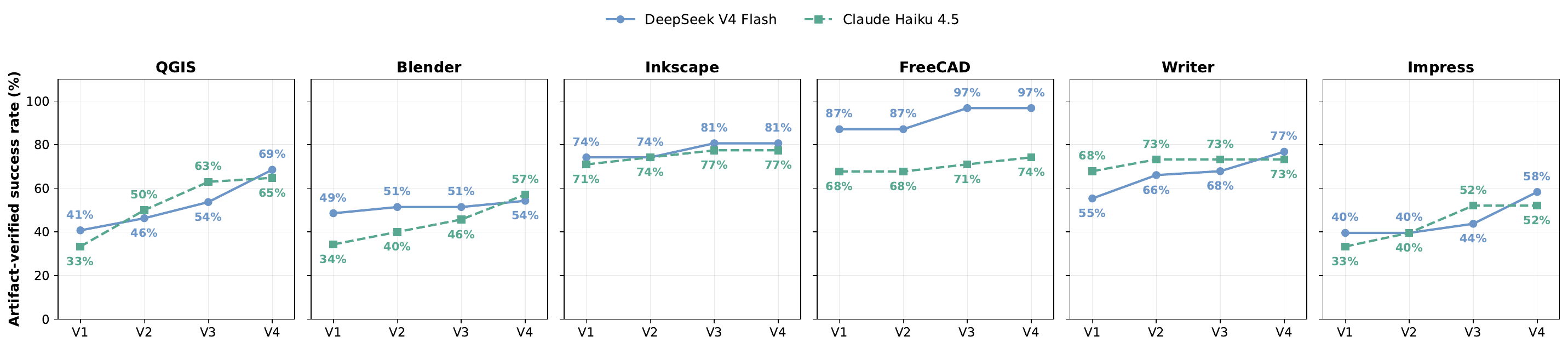}
\caption{Success rates across software releases (V1--V4: oldest to newest).}
\label{fig:versions}
\end{figure}

We observe consistent version trends across both backbones. 
Figure~\ref{fig:versions} shows the resulting success-rate trends for both models.
Although the two models differ in their overall capabilities, software-version changes produce similar performance patterns: the newest release outperforms the oldest in all model--application series.

\paragraph{Version gains persist across models.}
The newest release outperforms the oldest in all 12 model--application series. The gain is largest for QGIS, where DeepSeek rises from 41\% to 69\% and Haiku from 33\% to 65\%, close to 30 points for both models. For Impress, the corresponding changes are 40\% to 58\% and 33\% to 52\%. The magnitude varies by model and application, but the long-term direction is consistent. Although these releases were not developed specifically for LLM agents, improvements accumulated through ordinary software evolution can also make applications easier for agents to use. Section~\ref{sec:study-explanations} traces these gains to concrete interface changes.

\takeaway{Software evolution for human users can also improve agent performance: across two different backbones, the newest release consistently outperforms the oldest.}

\subsection{Cross-Version Performance Analysis}
\label{sec:study-explanations}
We further examine the agent trajectories to explain the observed gains. We divide the analysis into two parts: the first two interface properties are revealed through cross-version comparison, whereas the final property is a persistent limitation shared by all evaluated releases.

\subsubsection{Properties Revealed by Cross-Version Comparison}

\paragraph{Discoverability.}
The interface makes supported operations discoverable at runtime and describes their semantics and arguments clearly enough for agents to understand how to invoke them.

We compare the agent budget spent on SDK discovery and operation composition across versions as shown in Table~\ref{tab:version-interface}(a).
We find that the agent generally spends less time on SDK discovery when the interface upgrades, which is consistent with the cross-version accuracy gains in Section~\ref{sec:study-versions}.
Newer interfaces improve discoverability in two ways. First, higher cohesion makes capabilities easier to locate. In FreeCAD 0.21, reusable model parameters must be encoded indirectly as aliased spreadsheet cells. FreeCAD 1.0 introduces \code{App::\brk VarSet}, making them explicit properties of a dedicated object and therefore visible in the object model itself. Second, richer descriptions make discovered operations easier to use. From QGIS 3.10 to 3.44, mean runtime descriptions grow by 20\% per function, with more parameter semantics, behavioral notes, and examples. Together, these changes reduce the need to search elsewhere or probe calls blindly.

\paragraph{Granularity.}
The interface provides operations that closely match recurring user goals, reducing the need to compose low-level calls. A user goal often requires a long chain of low-level SDK calls, and each call is another point of failure and another block of tokens to spend. 

A release can shorten this chain by adding operations that sit closer to the user goal. QGIS's Processing toolbox grows from 272 to 354, 373, and 440 operations. Many are higher-level, ready-to-use processing operations, such as the geometry example below, directly supporting 22 of the 54 QGIS tasks in our benchmark. For Haiku, the mean QGIS chain length falls from 30.4 to 21.2 operations (Table~\ref{tab:version-interface}(b)). For example, geometry checks that once required manually iterating over features and vertices collapse into one call to \code{native:\brk check\brk geometry\brk missing\brk vertex}. Inkscape shows the same effect. For example, its multi-page document object lets agents express a multi-page export as a document-level operation instead of manually maintaining pages and composing page-by-page export calls, and the same task falls from 101 turns and 1.8M input tokens to 30 turns and 119k.

\begin{table}[!htbp]
\centering
\caption{Exploration turns (a) and SDK ops per task (b), normalized to each application's oldest release. $\blacktriangledown$/$\blacktriangle$: decrease/increase at the newest release; bold: change $\geq$20\%.}
\label{tab:version-interface}
\setlength{\tabcolsep}{4.2pt}
\renewcommand{\arraystretch}{1.12}
\resizebox{\linewidth}{!}{%
\begin{tabular}{ll *{6}{c}}
\toprule
\textbf{Metric} & \textbf{Model} & \textbf{QGIS} & \textbf{Blender} & \textbf{Inkscape} & \textbf{FreeCAD} & \textbf{Writer} & \textbf{Impress} \\
\midrule
\textit{(a) Exploration turns} & Haiku
 & \cNo{1.00}{1.01}{1.00} & \cDs{0.33}{0.14}{0.16} & \cDl{0.96}{0.97}{0.97}
 & \cUl{0.94}{1.06}{1.03} & \cDl{0.89}{0.90}{0.88} & \cUl{0.97}{0.63}{1.05} \\
 & DeepSeek
 & \cDm{0.55}{0.71}{0.69} & \cDs{0.82}{0.18}{0.32} & \cDl{0.92}{0.90}{0.92}
 & \cDm{0.91}{0.95}{0.76} & \cUm{1.06}{0.94}{1.28} & \cDm{0.94}{1.06}{0.75} \\
\addlinespace[2pt]\midrule
\textit{(b) SDK ops/task} & Haiku
 & \cDm{0.70}{0.71}{0.70} & \cNo{1.00}{0.95}{1.00} & \cDl{0.94}{0.92}{0.92}
 & \cNo{0.96}{0.89}{1.01} & \cDl{0.89}{1.02}{0.95} & \cDl{1.05}{0.87}{0.91} \\
 & DeepSeek
 & \cUl{0.95}{1.03}{1.04} & \cDm{0.88}{0.72}{0.80} & \cDl{0.70}{1.07}{0.86}
 & \cDl{0.94}{0.92}{0.96} & \cUl{1.19}{1.23}{1.06} & \cDl{1.01}{1.04}{0.92} \\
\bottomrule
\end{tabular}%
}
\end{table}
\subsubsection{Property Limited across Releases}

\paragraph{Observability.}
The interface exposes the resulting artifact state at the same level at which it is modified, so agents can verify whether the requested outcome has been achieved.
The two properties above emerge from cross-version changes: they vary across releases in ways that track changes in agent success and cost. Trajectory inspection also reveals failures of result verification. An audit restricted to the six study applications identifies 581 trajectories in which the agent reported completion but the saved artifact failed verification. Successful execution, as perceived by the agent, therefore does not ensure that the resulting artifact satisfies the task.

Of these 581 trajectories, 569 contain a detected check. Among these 569 trajectories, 153 (26.9\%) lack independent artifact readback and 416 (73.1\%) do. These counts distinguish observed checking behaviors but do not attribute failures to agent reasoning or unavailable state. Nevertheless, an interface can limit verification when it allows direct state modification without equally direct inspection of the result.

QGIS provides concrete examples. \code{QgsFillSymbol.\brk setAngle} directly sets a symbol angle, but no corresponding \code{angle()} reader exists. Recovering the value requires descending into individual symbol layers. More strongly, \code{QgsFeatureRenderer.\brk set\brk Vertex\brk Marker\brk Appearance} has no corresponding reader at all across the four evaluated releases. These cases show that state can be easy to modify yet difficult to inspect at the same abstraction level, limiting the agent's ability to verify its changes.

\takeaway{Agent-friendly interface quality has multiple dimensions. Discoverability and granularity improve in several newer releases and track gains in success and efficiency, while observability remains limited across the applications.}

\FloatBarrier

\section{Synthesizing Agent-Native Interfaces}
\label{sec:method}

Section~\ref{sec:study} measured what existing interfaces do to agents. We now state what an interface should do for agents. 
Section~\ref{sec:study-explanations} motivates three properties of agent-friendly interfaces:
\begin{itemize}[leftmargin=*,itemindent=0pt,labelsep=0.35em,topsep=1pt,itemsep=0pt,parsep=0pt]

\item \textbf{Discoverability.} The interface makes supported operations discoverable at runtime and describes their semantics and arguments clearly enough for agents to invoke them. Agents spend 24.1--36.2\% of input tokens on discovery (Figure~\ref{fig:overall-expense}), shipping documentation does not help (Appendix~\ref{sec:appendix-model-performance}).
\item \textbf{Granularity.} The interface provides operations that closely match recurring user goals, reducing the need to compose low-level calls (Table~\ref{tab:version-interface} (b)).
\item \textbf{Observability.} The interface exposes the resulting artifact state at the same level at which it is modified, so agents can verify whether the requested outcome has been achieved (Section~\ref{sec:study-explanations}).
\end{itemize}

Software evolution delivered the first two properties slowly and unevenly, over years of releases (Section~\ref{sec:study-explanations}), and never closed the gap in the third: write paths without matching read paths persist across every evaluated version. Our goal is to deliver all three for any installed application through automated synthesis and run-time adaptation, without waiting for the application to evolve.

\begin{figure}[t]
\centering
\includegraphics[width=\linewidth]{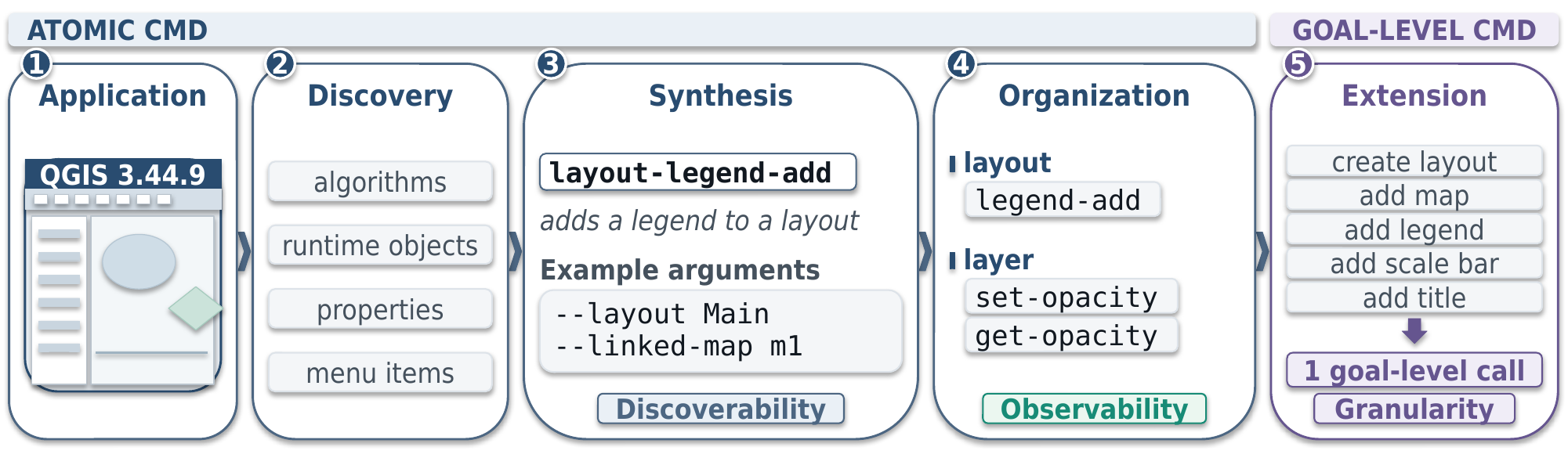}
\caption{Overview of \methodname{} with a QGIS example.}
\label{fig:method-overview}
\end{figure}

To this end, we design \methodname{} (Agent-Native Interface Synthesizer), which restructures the capabilities of an installed desktop application into an agent-oriented interface (Figure~\ref{fig:method-overview}). \methodname{} proceeds in two phases. At build time, it discovers capabilities from the running application, synthesizes each admitted capability into an \emph{atomic command} whose implementation has been validated by the application itself, and organizes the resulting commands so that agents can locate them without external documentation (discoverability) and read back the states they change when a read accessor exists (observability) (Section~\ref{sec:method-atomic}). Granularity cannot be settled at build time, because the goals worth promoting to higher-level operations become apparent only through use. At run time, \methodname{} therefore packages recurring command compositions into verified \emph{goal-level commands}, progressively aligning the interface with user goals (granularity) (Section~\ref{sec:method-goal}).
Together, the two phases deliver all three properties without changes to the underlying application. Appendix~\ref{sec:appendix-generation-time} reports interface generation times for the eight applications.

\subsection{Atomic Commands} \label{sec:method-atomic} 
An atomic command is the smallest unit of the interface \methodname{} synthesizes.
\methodname{} builds atomic commands in two stages with complementary goals. 
\emph{Capability discovery} enumerates what the installed application exposes and is tuned for recall: it may over-collect. \emph{Command synthesis} turns each candidate into an executable command and is tuned for precision: it releases only commands the application has demonstrably accepted.

\paragraph{Capability Discovery.}

To comprehensively recover the capabilities of an installed application, \methodname{} grounds discovery in the running application rather than in external documentation, which in our experiments provides no consistent gain in task success while increasing exploration overhead (Appendices~\ref{sec:appendix-model-performance} and~\ref{sec:appendix-token-distribution}). It draws on three complementary sources: registered operations such as QGIS processing algorithms and Blender operators, introspection of constructible objects and services for their methods and writable properties, and application menus for user-facing actions. These sources overlap but are individually incomplete, so their union forms the initial capability set. 
In implementation, a lightweight \emph{collector} shared across applications queries the exposed runtime interfaces, with an LLM formulating the application-specific queries, and normalizes the results into a common capability representation.

Not every discovered capability becomes an atomic command. 
\methodname{} retains capabilities that persistently modify the artifact or expose its state for inspection, filtering out those concerning only runtime machinery or transient application state. 
It makes this distinction by executing candidates against the running application rather than inferring it from names or documentation.

\paragraph{Command Synthesis.}
For each admitted capability, \methodname{} synthesizes an atomic command with a uniform interface. Each command pairs a specification of its arguments, valid values, and usage example with an implementation invoking the corresponding operation in the running application, as illustrated by the QGIS command in Figure~\ref{fig:method-overview}. Atomic commands take three forms: reading a property, writing a property, or invoking an operation on an artifact object. For each writable property, \methodname{} searches for a corresponding read accessor and, when one is available, exposes it as a read command so that the agent can inspect the resulting state.

The implementation of a command is derived from the runtime evidence collected during capability discovery: the parameters, types, and defaults of the corresponding operation or property as reported by the running application. 
An LLM uses this runtime information to construct a candidate invocation, which is then executed against the application. 
A successful return only establishes that the application accepted the invocation, not that the resulting state persists in the saved artifact, so \methodname{} retains only invocations whose effects can be observed after saving and reopening the artifact. 
Each released command is thus bound to an invocation already validated by the installed application, without assuming prior application-specific API knowledge.

This process can produce thousands of atomic commands, so exposing them as a flat list would make discovery expensive again. \methodname{} therefore organizes commands by the artifact objects they operate on and names each command by its object and action. Command browsing details are provided in Appendix~\ref{sec:appendix-command-organization}.
Across the eight applications, interface generation takes 17.1--36.4 hours per application in wall-clock time (Appendix~\ref{sec:appendix-generation-time}).

\subsection{Goal-Level Commands}
\label{sec:method-goal}

To improve granularity, \methodname{} augments atomic commands with \emph{goal-level commands} that capture recurring user goals as higher-level operations. Rather than fixing this layer in advance, \methodname{} synthesizes goal-level commands on demand as requests accumulate at runtime. In our experiments, this stream is simulated with agent trajectories from previously solved tasks, and Appendix~\ref{sec:appendix-goal-holdout} confirms the gain on a disjoint held-out split. Each released goal-level command is a verified composition of atomic commands that produces the intended artifact state.

\paragraph{Where Goals Come From.}
Enumerating possible compositions is impractical and mostly yields sequences no user needs. 
\methodname{} instead derives candidates from actual use. As agents serve requests, it records the atomic-command sequences they execute and counts subsequences that recur across requests. 
Recurrence alone does not imply a user goal, since the same implementation pattern may be shared by otherwise unrelated requests, so it serves only to surface potentially reusable compositions. 
Once a subsequence recurs in at least three requests, \methodname{} examines the requests that produced it to determine whether the sequence realizes a common artifact-level subgoal across them. 
Here, a goal is a shared artifact-level subgoal, not necessarily the complete user request.

\paragraph{From Sequence to Command.}
A recurring sequence has no user-facing identity of its own, so \methodname{} turns it into a named command only when such a shared goal is found. 
Command naming, parameterization, and execution details are provided in Appendix~\ref{sec:appendix-command-packaging}. 
Before release, \methodname{} performs a reproducibility check by rerunning the command on a clean copy of the originating artifact and verifying the saved final state. The released command is then exposed through the same interface and invoked like any atomic command.

\subsection{Performance Evaluation}
\label{sec:method-comparison}

\subsubsection{Experiment Settings}

We reuse the study's scaffold, OpenCode~\citep{opencode2026}, and both of its backbones, Claude Haiku 4.5 and DeepSeek V4 Flash, following the success criterion in Section~\ref{sec:study-setup}. We use all 255 \sysname{} tasks at their newest release, plus 76 Calc and 37 GIMP tasks collected with the same pipeline. Calc and GIMP were held out from every stage of method development: their interfaces were synthesized once, after the pipeline was frozen, and never inspected before evaluation.

We compare \methodname{} against two interfaces: the application's native SDK, accessed as in Section~\ref{sec:study-setup}, and the interface generated by CLI-Anything~\citep{yang2026clianything}, obtained by installing its plugin in Claude Code. Both generated interfaces are constructed with Claude Code~\citep{anthropic2025claudecode} using Claude Opus 4.8~\citep{anthropic2026opus48}.

\subsubsection{Experiment Results}

\begin{table}[!htbp]
\centering
\setlength{\tabcolsep}{3.2pt}
\renewcommand{\arraystretch}{1.12}
\caption{Success rate (\%) across eight applications with DeepSeek V4 Flash as the backbone.}
\resizebox{\linewidth}{!}{
\begin{tabular}{lrrrrrrrrrr}
\toprule
& \multicolumn{8}{c}{\textbf{Success rate (\%)}} & \multicolumn{2}{c}{\textbf{Overall}} \\
\cmidrule(lr){2-9}\cmidrule(lr){10-11}
\textbf{Interface} & \textbf{QGIS} & \textbf{Writer} &
\textbf{Blender} & \textbf{GIMP} & \textbf{Calc} & \textbf{Impress} &
\textbf{Inkscape} & \textbf{FreeCAD} & \textbf{Success (\%)} & \textbf{Tokens per task} \\
\midrule
Native SDK & 68.5 & 76.8 & 54.3 & 67.6 & 59.2 & 58.3 & 80.6 & 96.8 & 68.5 & 84.4K \\
CLI-Anything & 13.0 & 10.7 & 8.6 & 43.2 & 6.6 & 2.1 & 38.7 & 45.2 & 17.4 & 67.0K (20.7\%$\downarrow$) \\
\methodname{} w/o goal-level & 77.8 & 89.3 & 45.7 & 75.7 & 63.2 & 66.7 & 83.9 & 77.4 & 72.3 & 22.6K (73.2\%$\downarrow$) \\
\methodname{} w/ goal-level & 81.5 & 89.3 & 51.4 & 75.7 & 68.4 & 66.7 & 83.9 & 77.4 & 74.5 & 21.2K (74.9\%$\downarrow$) \\
\bottomrule
\end{tabular}
}
\label{tab:method-cross-app}
\end{table}

\paragraph{Overall Performance.}
Atomic commands raise overall success from 68.5\% to 72.3\%, and adding goal-level commands raises it to 74.5\%, though both remain below the native SDK on FreeCAD and Blender (Table~\ref{tab:method-cross-app}). Tokens per task fall from 84.4K with the native SDK to 22.6K and 21.2K, respectively. These savings reflect easier command discovery and the packaging of recurring operations into goal-level commands, reducing exploration and multi-step composition as illustrated below. With Claude Haiku 4.5 as the backbone, the ordering of the interfaces is unchanged (Appendix~\ref{sec:appendix-haiku}).

\paragraph{Capability Coverage.}
Coverage explains both the gap to CLI-Anything and the two exceptions above.
Across the applications covered by both systems, \methodname{} exposes over $20\times$ as many commands as CLI-Anything. CLI-Anything's interface omits direct access to some application capabilities, such as QGIS print-layout creation.
Of CLI-Anything's failed tasks, 98.7\% are attributable to missing commands and only 1.3\% to agent errors with the necessary commands available, so its 17.4\% success rate reflects what the interface omits rather than the command-line form itself.

\methodname{}'s remaining gap is of a different kind: discovery is exhaustive, but validation can reject operations whose preconditions are implicit. FreeCAD operations depend on linked sketches and modeling features, and Blender operations on coordinated scene objects, selections, and editing modes. Such dependencies are harder to reproduce automatically than explicit-input operations such as QGIS processing algorithms, so a small set of capabilities available through the native SDK remains uncovered (28.6\% of \methodname{}'s FreeCAD failures lack a required command). This gap occurs only in the two applications with the most implicit object relationships.

\paragraph{Observation Command Ablation.} 
 We test the contribution of observation commands directly by prohibiting observation commands through agent instructions while leaving all editing commands available.  
As reported in Appendix~\ref{sec:appendix-observability}, prohibiting observation commands by instruction reduces overall success from 72.3\% to 60.3\%, a drop of 12.0 percentage points. 

\paragraph{Goal-Level Commands.}
Goal-level commands target the granularity property. We synthesize them from trajectories of tasks solved with atomic commands, yielding 10 commands across the eight applications, and evaluate the augmented interface on the full task set. These commands package recurring multi-step sequences into single invocations. For example, five QGIS operations to create a print layout and add its map, legend, scale bar, and title become one parameterized call.

Adding goal-level commands never lowers success on any application and raises the overall rate from 72.3\% to 74.5\%. Further analysis is provided in Appendix~\ref{sec:appendix-goal-full}.

\FloatBarrier

\section{Conclusion}

LLM agents increasingly operate desktop applications through interfaces designed for humans and developers. Using software evolution as natural interface variation, we introduce \sysname, 255 artifact-verified tasks across six applications, and identify three properties that shape agent performance: discoverability, granularity, and observability. 
Evolution delivered the first two slowly and unevenly, and never closed the gap in the third. \methodname{} delivers all three for any installed application, without waiting for it to evolve, by synthesizing its runtime capabilities into commands validated by the application itself and extending them with goal-level commands mined from use. 
Across eight applications, it raises success from 68.5\% with the native SDK to 74.5\%, while removing observation commands alone costs 12.0 points. The interface an agent is given is thus a first-class design choice, and one that can be synthesized rather than waited for.

\subsection*{AI Use Statement}

We used generative AI tools (LLMs) in preparing this paper for four purposes: polishing the grammar and clarity of author-written text, finding candidate related work, implementing experiment and analysis code, and synthesizing benchmark tasks. LLMs are also part of the studied systems: the evaluated agents are LLMs (Section~\ref{sec:study-setup}), \methodname{} relies on LLM coding agents to synthesize commands, and both generated interfaces in Section~\ref{sec:method-comparison} are built with Claude Code~\citep{anthropic2025claudecode} using Claude Opus 4.8. The authors checked every edit against the intended meaning, reviewed and tested all AI-assisted code, checked every retrieved reference against its original source, and accepted a synthesized task only after its verifier accepted the reference artifact and rejected task-relevant perturbations (Section~\ref{sec:dataset-tasks}). The authors take full responsibility for the content of this paper.

\subsection*{Reproducibility Statement}

The code, \sysname{} tasks with their verifiers, and evaluation scripts are available at \url{https://github.com/co-evolve-lab/ANIS}.

\FloatBarrier
\bibliography{custom}

\begin{thebibliography}{34}
\providecommand{\natexlab}[1]{#1}
\providecommand{\url}[1]{\texttt{#1}}
\expandafter\ifx\csname urlstyle\endcsname\relax
  \providecommand{\doi}[1]{doi: #1}\else
  \providecommand{\doi}{doi: \begingroup \urlstyle{rm}\Url}\fi

\bibitem[Bonatti et~al.(2025)Bonatti, Zhao, Bonacci, Dupont, Abdali, Li, Lu, Wagle, Koishida, Bucker, Jang, and Hui]{bonatti2025windows}
Rogerio Bonatti, Dan Zhao, Francesco Bonacci, Dillon Dupont, Sara Abdali, Yinheng Li, Yadong Lu, Justin Wagle, Kazuhito Koishida, Arthur Bucker, Lawrence~Keunho Jang, and Zheng Hui.
\newblock {Windows Agent Arena}: Evaluating multi-modal {OS} agents at scale.
\newblock In \emph{International Conference on Machine Learning}, 2025.

\bibitem[Nayak et~al.(2025)Nayak, Jian, Lin, Rodr{\'{\i}}guez, Kalsi, Awal, Chapados, {\"{O}}zsu, Agrawal, V{\'{a}}zquez, Pal, Taslakian, Gella, and Rajeswar]{DBLP:journals/corr/abs-2503-15661}
Shravan Nayak, Xiangru Jian, Kevin~Qinghong Lin, Juan~A. Rodr{\'{\i}}guez, Montek Kalsi, Rabiul Awal, Nicolas Chapados, M.~Tamer {\"{O}}zsu, Aishwarya Agrawal, David V{\'{a}}zquez, Christopher Pal, Perouz Taslakian, Spandana Gella, and Sai Rajeswar.
\newblock {UI-Vision}: A desktop-centric {GUI} benchmark for visual perception and interaction.
\newblock In \emph{International Conference on Machine Learning}, 2025.

\bibitem[Zhao et~al.(2025)Zhao, Yang, Yu, Gao, and Shou]{DBLP:journals/corr/abs-2502-08047}
Henry~Hengyuan Zhao, Kaiming Yang, Wendi Yu, Difei Gao, and Mike~Zheng Shou.
\newblock {WorldGUI}: An interactive benchmark for desktop {GUI} automation from any starting point.
\newblock \emph{arXiv preprint arXiv:2502.08047}, 2025.

\bibitem[Li et~al.(2026)Li, Guo, Wang, Guo, Tao, Shan, Liu, Chen, Song, Tang, Zhang, and Zheng]{li2026advances}
Caihua Li, Lianghong Guo, Yanlin Wang, Daya Guo, Wei Tao, Zhenyu Shan, Mingwei Liu, Jiachi Chen, Haoyu Song, Duyu Tang, Hongyu Zhang, and Zibin Zheng.
\newblock Advances and frontiers of llm-based issue resolution in software engineering: A comprehensive survey.
\newblock \emph{arXiv preprint arXiv:2601.11655}, 2026.

\bibitem[Lu et~al.(2025)Lu, Zhang, Yang, Zhang, Wang, Du, Lin, Rajmohan, Zhang, and Zhang]{lu2025axis}
Junting Lu, Zhiyang Zhang, Fangkai Yang, Jue Zhang, Lu~Wang, Chao Du, Qingwei Lin, Saravan Rajmohan, Dongmei Zhang, and Qi~Zhang.
\newblock {AXIS}: Efficient human-agent-computer interaction with {API}-first {LLM}-based agents.
\newblock In \emph{Annual Meeting of the Association for Computational Linguistics}, 2025.

\bibitem[Song et~al.(2025)Song, Xu, Zhou, and Neubig]{song2025beyondbrowsing}
Yueqi Song, Frank~F. Xu, Shuyan Zhou, and Graham Neubig.
\newblock Beyond browsing: {API}-based web agents.
\newblock In \emph{Findings of the Association for Computational Linguistics: ACL 2025}, 2025.

\bibitem[De~Masi(2026)]{demasi2026terminal}
Alexandre De~Masi.
\newblock Terminal is all you need: Design properties for human--{AI} agent collaboration.
\newblock \emph{arXiv preprint arXiv:2603.10664}, 2026.

\bibitem[Yang et~al.(2026)Yang, Fan, and Huang]{yang2026clianything}
Yuhao Yang, Tianyu Fan, and Chao Huang.
\newblock {CLI-Anything}: Towards agent-native computer use.
\newblock \emph{arXiv preprint arXiv:2606.03854}, 2026.

\bibitem[Wang et~al.(2026)Wang, Li, and Chen]{wang2026declarative}
Yuan Wang, Mingyu Li, and Haibo Chen.
\newblock From imperative to declarative: Towards {LLM}-friendly {OS} interfaces for boosted computer-use agents.
\newblock In \emph{European Conference on Computer Systems}, 2026.

\bibitem[Zhou et~al.(2026)Zhou, Zhang, Zhao, Wei, Song, Cohan, and Zhao]{zhou2026guivscli}
Xiao Zhou, Siyue Zhang, Yilun Zhao, Jinbiao Wei, Tingyu Song, Arman Cohan, and Chen Zhao.
\newblock {GUI} vs. {CLI}: Execution bottlenecks in screen-only and skill-mediated computer-use agents.
\newblock \emph{arXiv preprint arXiv:2606.24551}, 2026.

\bibitem[{QGIS Development Team}(2026)]{qgis2026}
{QGIS Development Team}.
\newblock {QGIS Geographic Information System}, 2026.
\newblock URL \url{https://qgis.org/}.

\bibitem[Xie et~al.(2024)Xie, Zhang, Chen, Li, Zhao, Cao, Hua, Cheng, Shin, Lei, Liu, Xu, Zhou, Savarese, Xiong, Zhong, and Yu]{xie2024osworld}
Tianbao Xie, Danyang Zhang, Jixuan Chen, Xiaochuan Li, Siheng Zhao, Ruisheng Cao, Toh~Jing Hua, Zhoujun Cheng, Dongchan Shin, Fangyu Lei, Yitao Liu, Yiheng Xu, Shuyan Zhou, Silvio Savarese, Caiming Xiong, Victor Zhong, and Tao Yu.
\newblock {OSWorld}: Benchmarking multimodal agents for open-ended tasks in real computer environments.
\newblock In \emph{Advances in Neural Information Processing Systems}, 2024.

\bibitem[Agashe et~al.(2025)Agashe, Han, Gan, Yang, Li, and Wang]{agashe2025agents}
Saaket Agashe, Jiuzhou Han, Shuyu Gan, Jiachen Yang, Ang Li, and Xin~Eric Wang.
\newblock {Agent S}: An open agentic framework that uses computers like a human.
\newblock In \emph{International Conference on Learning Representations}, 2025.

\bibitem[Wang et~al.(2025)Wang, Xu, Wang, Zhang, Yan, Zhang, Huang, and Ji]{wang2025mobileagente}
Zhenhailong Wang, Haiyang Xu, Junyang Wang, Xi~Zhang, Ming Yan, Ji~Zhang, Fei Huang, and Heng Ji.
\newblock {Mobile-Agent-E}: Self-evolving mobile assistant for complex tasks.
\newblock \emph{arXiv preprint arXiv:2501.11733}, 2025.

\bibitem[Jiang et~al.(2025)Jiang, Zhuang, Song, Yang, Zhou, and Zhang]{jiang2025appagentx}
Wenjia Jiang, Yangyang Zhuang, Chenxi Song, Xu~Yang, Joey~Tianyi Zhou, and Chi Zhang.
\newblock {AppAgentX}: Evolving {GUI} agents as proficient smartphone users.
\newblock \emph{arXiv preprint arXiv:2503.02268}, 2025.

\bibitem[Zhang et~al.(2026)Zhang, Huang, Ni, Mu, Qin, He, Wang, Yang, Zhao, Qiao, Du, Li, Kang, Jiang, Zheng, Wang, Qian, Ma, Lou, Lin, Rajmohan, and Zhang]{zhang2025ufo2}
Chaoyun Zhang, He~Huang, Chiming Ni, Jian Mu, Si~Qin, Shilin He, Lu~Wang, Fangkai Yang, Pu~Zhao, Bo~Qiao, Chao Du, Liqun Li, Yu~Kang, Zhao Jiang, Suzhen Zheng, Rujia Wang, Jiaxu Qian, Minghua Ma, Jian-Guang Lou, Qingwei Lin, Saravan Rajmohan, and Dongmei Zhang.
\newblock {UFO2}: The desktop {AgentOS}.
\newblock \emph{Transactions on Machine Learning Research}, 2026.

\bibitem[Zhao et~al.(2026)Zhao, Liu, Liang, He, Lu, Wang, Huang, Chai, Kang, Guo, Wang, Zhang, Liu, and Liu]{zhao2025masbench}
Pengxiang Zhao, Guangyi Liu, Yaozhen Liang, Weiqing He, Zhengxi Lu, WenHao Wang, Yuehao Huang, Yuxiang Chai, Zhaolu Kang, Yaxuan Guo, Hao Wang, Kexin Zhang, Liang Liu, and Yong Liu.
\newblock {MAS-Bench}: A unified benchmark for shortcut-augmented hybrid mobile {GUI} agents.
\newblock In \emph{Annual Meeting of the Association for Computational Linguistics}, 2026.

\bibitem[Zhang et~al.(2025)Zhang, He, Li, Qin, Kang, Lin, Rajmohan, and Zhang]{zhang2025apigui}
Chaoyun Zhang, Shilin He, Liqun Li, Si~Qin, Yu~Kang, Qingwei Lin, Saravan Rajmohan, and Dongmei Zhang.
\newblock {API} agents vs. {GUI} agents: Divergence and convergence.
\newblock \emph{arXiv preprint arXiv:2503.11069}, 2025.

\bibitem[Ni et~al.(2025)Ni, Jian, Wang, Shah, and Hong]{ni2025doc2agent}
Xinyi Ni, Haonan Jian, Qiuyang Wang, Vedanshi~Chetan Shah, and Pengyu Hong.
\newblock {Doc2Agent}: Scalable generation of tool-using agents from {API} documentation.
\newblock \emph{arXiv preprint arXiv:2506.19998}, 2025.

\bibitem[Agarwal et~al.(2025)Agarwal, Gupta, Soni, Vallam, Sindhgatta, and Mehta]{agarwal_automated_2025}
Prerna Agarwal, Himanshu Gupta, Soujanya Soni, Rohith Vallam, Renuka Sindhgatta, and Sameep Mehta.
\newblock Automated creation and enrichment framework for improved invocation of enterprise {APIs} as tools.
\newblock \emph{arXiv preprint arXiv:2509.11626}, 2025.

\bibitem[Bandlamudi et~al.(2025)Bandlamudi, Chaudhuri, Gantayat, Ghosh, Mukherjee, Agarwal, Sindhgatta, and Mehta]{bandlamudi_framework_2025}
Jayachandu Bandlamudi, Ritwik Chaudhuri, Neelamadhav Gantayat, Sambit Ghosh, Kushal Mukherjee, Prerna Agarwal, Renuka Sindhgatta, and Sameep Mehta.
\newblock A framework for testing and adapting {REST} {APIs} as {LLM} tools.
\newblock \emph{arXiv preprint arXiv:2504.15546}, 2025.

\bibitem[Sigdel and Baral(2026)]{sigdel_schema_2026}
Akshey Sigdel and Rista Baral.
\newblock Schema first tool {APIs} for {LLM} agents: A controlled study of tool misuse, recovery, and budgeted performance.
\newblock \emph{arXiv preprint arXiv:2603.13404}, 2026.

\bibitem[Sadani and Kumar(2026)]{sadani2026toolattention}
Anuj Sadani and Deepak Kumar.
\newblock Tool attention is all you need: Dynamic tool gating and lazy schema loading for eliminating the {MCP}/tools tax in scalable agentic workflows.
\newblock \emph{arXiv preprint arXiv:2604.21816}, 2026.

\bibitem[Wei et~al.(2026)Wei, Ma, Zhao, Zhou, Ni, Gan, and Cohan]{opencomputer2026}
Jinbiao Wei, Qianran Ma, Yilun Zhao, Xiao Zhou, Kangqi Ni, Guo Gan, and Arman Cohan.
\newblock {OpenComputer}: Verifiable software worlds for computer-use agents.
\newblock \emph{arXiv preprint arXiv:2605.19769}, 2026.

\bibitem[Zheng et~al.(2026)Zheng, Zou, Ye, Liu, Kong, Zhang, Zheng, and Pan]{zheng2026ebench}
Weihuang Zheng, Tianyuan Zou, Eileen Ye, Alphet Liu, Youyong Kong, Ya-Qin Zhang, Duran Zheng, and Maxm Pan.
\newblock {E-Bench}: Benchmarking multi-step tool-use agents in real-world product scenarios.
\newblock \emph{arXiv preprint arXiv:2607.23722}, 2026.

\bibitem[Vats and Golev(2026)]{vats2026scaffold}
Naman Vats and Oleg Golev.
\newblock The scaffold effect in coding agents: Harness choice as a hidden variable in coding-agent evaluation.
\newblock \emph{arXiv preprint arXiv:2607.22585}, 2026.

\bibitem[Rombaut(2026)]{rombaut2026scaffold}
Benjamin Rombaut.
\newblock Inside the scaffold: A source-code taxonomy of coding agent architectures.
\newblock \emph{arXiv preprint arXiv:2604.03515}, 2026.

\bibitem[{OpenCode Contributors}(2026)]{opencode2026}
{OpenCode Contributors}.
\newblock {OpenCode}: The open source {AI} coding agent, 2026.
\newblock URL \url{https://github.com/anomalyco/opencode}.
\newblock Version 1.17.14.

\bibitem[{Anthropic}(2025{\natexlab{a}})]{anthropic2025haiku45}
{Anthropic}.
\newblock Introducing {Claude Haiku 4.5}, 2025{\natexlab{a}}.
\newblock URL \url{https://www.anthropic.com/news/claude-haiku-4-5}.

\bibitem[{DeepSeek-AI}(2026)]{deepseek2026v4}
{DeepSeek-AI}.
\newblock {DeepSeek-V4}: Towards highly efficient million-token context intelligence, 2026.
\newblock URL \url{https://huggingface.co/deepseek-ai/DeepSeek-V4-Flash}.

\bibitem[{Anthropic}(2025{\natexlab{b}})]{anthropic2025claudecode}
{Anthropic}.
\newblock Claude code.
\newblock \url{https://www.anthropic.com/claude-code}, 2025{\natexlab{b}}.
\newblock Agentic coding tool.

\bibitem[{Anthropic}(2026{\natexlab{a}})]{anthropic2026opus48}
{Anthropic}.
\newblock Introducing {Claude Opus 4.8}, 2026{\natexlab{a}}.
\newblock URL \url{https://www.anthropic.com/news/claude-opus-4-8}.

\bibitem[{Kimi Team}(2026)]{kimiteam2026k25}
{Kimi Team}.
\newblock {Kimi K2.5}: Visual agentic intelligence.
\newblock \emph{arXiv preprint arXiv:2602.02276}, 2026.

\bibitem[{Anthropic}(2026{\natexlab{b}})]{anthropic2026opus46}
{Anthropic}.
\newblock Introducing {Claude Opus 4.6}, 2026{\natexlab{b}}.
\newblock URL \url{https://www.anthropic.com/news/claude-opus-4-6}.

\end{thebibliography}
\bibliographystyle{unsrtnat}

\clearpage
\beginappendix
\section{Overall Model Performance and Documentation}
\label{sec:appendix-model-performance}

We compare four backbone models, Claude Haiku 4.5~\citep{anthropic2025haiku45}, Kimi K2.5~\citep{kimiteam2026k25}, DeepSeek V4 Flash~\citep{deepseek2026v4}, and Claude Opus 4.6~\citep{anthropic2026opus46}, with and without official documentation for all eight desktop applications. All runs use the OpenCode agent scaffold~\citep{opencode2026} to operate each application through its native SDK. For each model, both conditions use the same tasks and artifact verifiers. In the documentation condition, we provide offline copies of the official documentation, specify their local paths in the agent instructions, and require the agent to consult them when uncertain about the usage. In the other condition, no additional documentation is provided.
Table~\ref{tab:appendix-model-performance} reports the available success rates. 

\begin{table}[htbp]
\centering
\setlength{\tabcolsep}{3.5pt}
\renewcommand{\arraystretch}{1.12}
\caption{Success rate (\%) with and without official documentation. }
\resizebox{\linewidth}{!}{%
\begin{tabular}{lrrrrrrrrr}
\toprule
\textbf{Model} & \textbf{QGIS} & \textbf{Writer} & \textbf{Blender} & \textbf{GIMP} & \textbf{Calc} & \textbf{Impress} & \textbf{Inkscape} & \textbf{FreeCAD} & \textbf{Overall} \\
\midrule
\multicolumn{10}{l}{\cellcolor[rgb]{0.88,0.95,1}\textit{Without documentation}} \\
Claude Haiku 4.5  & 64.8 & 73.2 & 57.1 & 59.5 & 47.4 & 52.1 & 77.4 & 74.2 & 61.4 \\
Kimi K2.5        & 75.9 & 73.2 & 57.1 & 75.7 & 56.6 & 68.8 & 83.9 & 93.5 & 70.9 \\
DeepSeek V4 Flash & 68.5 & 76.8 & 54.3 & 67.6 & 59.2 & 58.3 & 80.6 & 96.8 & 68.5 \\
Claude Opus 4.6   & 77.8 & 87.5 & 71.4 & 78.4 & 61.8 & 77.1 & 87.1 & 93.5 & 77.4 \\
\midrule
\multicolumn{10}{l}{\cellcolor[rgb]{0.88,0.95,1}\textit{With documentation}} \\
Claude Haiku 4.5  & 61.1 & 67.9 & 48.6 & 64.9 & 48.7 & 54.2 & 77.4 & 74.2 & 60.3 \\
Kimi K2.5        & 77.8 & 73.2 & 57.1 & 75.7 & 57.9 & 75.0 & 83.9 & 93.5 & 72.3 \\
DeepSeek V4 Flash & 66.7 & 71.4 & 48.6 & 73.0 & 52.6 & 64.6 & 80.6 & 96.8 & 66.8 \\
Claude Opus 4.6   & 72.2 & 83.9 & 68.6 & 78.4 & 59.2 & 79.2 & 87.1 & 93.5 & 75.5 \\
\bottomrule
\end{tabular}%
}
\label{tab:appendix-model-performance}
\end{table}

Without documentation, overall success ranges from 61.4\% to 77.4\%. Providing documentation raises Kimi's success by 1.4 percentage points, while Haiku, DeepSeek, and Opus decline by 1.1, 1.7, and 1.9 points. Effects also vary within a model: Haiku improves on GIMP but declines on Blender. Documentation therefore provides no consistent gain across these backbones and applications.

\FloatBarrier
\section{Token Budget Distribution}
\label{sec:appendix-token-distribution}

\begin{figure}[htbp]
\centering
\includegraphics[width=0.90\linewidth]{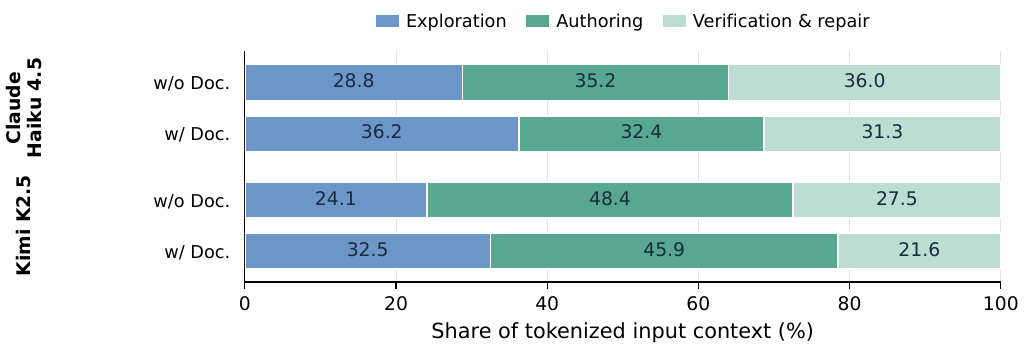}
\caption{Token budget distribution across activities for two models w/ and w/o Doc.}
\label{fig:overall-expense}
\end{figure}

We audit the trajectories of Claude Haiku 4.5 and Kimi K2.5 on all tasks. We attribute context tokens to three activities: \emph{exploration}, consulting references, probing interfaces, and inspecting task files; \emph{authoring}, planning, writing, and executing task code; and \emph{verification and repair}, checking outputs and responding to tool errors. Figure~\ref{fig:overall-expense} compares bundled references alone with an additional documentation library.

Adding the library and requiring consultation increases the exploration share from 28.8\% to 36.2\% for Haiku and from 24.1\% to 32.5\% for Kimi. However, documentation does not consistently improve overall success. Some models perform worse (Table~\ref{tab:appendix-model-performance}). Additional material may introduce irrelevant or version-mismatched information, while agents can still overlook key constraints or misuse the documented APIs.

The models also differ in how they allocate their tokens. Kimi places more weight on writing and executing code, with a smaller share devoted to verification and repair. Haiku distributes its budget more evenly between implementation and checking or repair.

\FloatBarrier

\section{Performance Evaluation with Claude Haiku 4.5}
\label{sec:appendix-haiku}

Table~\ref{tab:method-cross-app} reports the interface comparison with DeepSeek V4 Flash as the backbone. To check whether the effect of the interface depends on the backbone model, we repeat the comparison with Claude Haiku 4.5, keeping the remaining settings of Section~\ref{sec:method-comparison} fixed: the same scaffold, the same 368 tasks, and the same generated interfaces for CLI-Anything and \methodname{}. Table~\ref{tab:appendix-haiku} reports the results.

\begin{table}[htbp]
\centering
\setlength{\tabcolsep}{3.2pt}
\renewcommand{\arraystretch}{1.12}
\caption{Success rate (\%) across eight applications with Claude Haiku 4.5 as the backbone. \methodname{} is evaluated with atomic commands alone and with added goal-level commands. Token reductions are relative to the native SDK.}
\resizebox{\linewidth}{!}{
\begin{tabular}{lrrrrrrrrrr}
\toprule
& \multicolumn{8}{c}{\textbf{Success rate (\%)}} & \multicolumn{2}{c}{\textbf{Overall}} \\
\cmidrule(lr){2-9}\cmidrule(lr){10-11}
\textbf{Interface} & \textbf{QGIS} & \textbf{Writer} &
\textbf{Blender} & \textbf{GIMP} & \textbf{Calc} & \textbf{Impress} &
\textbf{Inkscape} & \textbf{FreeCAD} & \textbf{Success (\%)} & \textbf{Tokens per task} \\
\midrule
Native SDK & 64.8 & 73.2 & 57.1 & 59.5 & 47.4 & 52.1 & 77.4 & 74.2 & 61.4 & 85.0K \\
CLI-Anything & 13.0 & 10.7 & 8.6 & 37.8 & 5.3 & 2.1 & 38.7 & 35.5 & 15.8 & 67.0K (21.2\%$\downarrow$) \\
\methodname{} w/o goal-level & 74.1 & 83.9 & 48.6 & 67.6 & 50.0 & 60.4 & 80.6 & 58.1 & 64.9 & 23.0K (72.9\%$\downarrow$) \\
\methodname{} w/ goal-level & 79.6 & 83.9 & 54.3 & 67.6 & 55.3 & 60.4 & 80.6 & 58.1 & 67.4 & 22.0K (74.1\%$\downarrow$) \\
\bottomrule
\end{tabular}
}
\label{tab:appendix-haiku}
\end{table}

\paragraph{Results.}
The results mirror those with DeepSeek V4 Flash. \methodname{} raises overall success from 61.4\% to 64.9\% with atomic commands and to 67.4\% with goal-level commands, while reducing tokens per task by over 70\%, indicating that the gains stem from the interface rather than from a particular backbone model.

\FloatBarrier

\section{Observation Command Ablation}
\label{sec:appendix-observability}

Observability is the one property that software evolution never delivered (Section~\ref{sec:method}).

We assess the value of observation commands using DeepSeek V4 Flash with OpenCode, following the experimental settings in Section~\ref{sec:method-comparison}. We compare the full generated interface with a condition that prohibits observation commands, keeping all other settings fixed.

The ablation uses the atomic-command interface on all 368 tasks, with the unablated baseline rerun under the same setting for comparability.

In the ablated condition, the instructions explicitly identify the generated commands that inspect artifact state and prohibit their use. To verify compliance, we conduct an after-the-fact audit by extracting shell commands from the saved agent trajectories and checking them against the prohibited command set. The audit confirms that the model follows the restriction across all tasks, with no observation commands invoked.

\begin{table}[htbp]
\centering
\setlength{\tabcolsep}{4pt}
\renewcommand{\arraystretch}{1.12}
\caption{Success rate (\%) with observation commands allowed or prohibited by instruction.}
\resizebox{\linewidth}{!}{
\begin{tabular}{lrrrrrrrrr}
\toprule
\textbf{Condition} & \textbf{QGIS} & \textbf{Writer} & \textbf{Blender} & \textbf{GIMP} & \textbf{Calc} & \textbf{Impress} & \textbf{Inkscape} & \textbf{FreeCAD} & \textbf{Overall} \\
\midrule
Tasks & 54 & 56 & 35 & 37 & 76 & 48 & 31 & 31 & 368 \\
\midrule
Observation allowed & 77.8 & 89.3 & 45.7 & 75.7 & 63.2 & 66.7 & 83.9 & 77.4 & 72.3 \\
Observation prohibited & 72.2 & 69.6 & 31.4 & 75.7 & 32.9 & 66.7 & 83.9 & 71.0 & 60.3 \\
Change (pp) & $-5.6$ & $-19.7$ & $-14.3$ & $0.0$ & $-30.3$ & $0.0$ & $0.0$ & $-6.4$ & $-12.0$ \\
\bottomrule
\end{tabular}
}
\label{tab:appendix-observation-ablation}
\end{table}

\paragraph{Results.}
Table~\ref{tab:appendix-observation-ablation} shows that prohibiting observation reduces overall success from 72.3\% to 60.3\%, a drop of 12.0 percentage points. These results suggest that access to editing commands alone is often insufficient: agents also need artifact state to choose parameters and detect deviations from the requested outcome.

Calc and Writer show the largest decreases, at 30.3 and 19.7 points, respectively, while GIMP, Impress, and Inkscape show no aggregate change. These results suggest that access to editing commands alone is often insufficient: agents also need to inspect artifact state to choose appropriate actions and verify their effects, which is what the paired read accessors in Section~\ref{sec:method-atomic} provide.

\paragraph{Blender Example.}
In a task requiring a chair to face a table, the observation-enabled agent reads the initial orientation, sets the chair's rotation accordingly, and verifies the saved scene, passing the task. Without observation, the agent applies a $180^\circ$ relative rotation and receives a successful command response, but the chair remains perpendicular to the required direction and fails verification. Command success therefore does not establish task success: reading the initial state and checking the resulting orientation are needed to connect an edit to the intended spatial relation.

\FloatBarrier

\section{Goal-Level Command Evaluation}
\label{sec:appendix-goal-evaluation}

\subsection{Full-Task Evaluation}
\label{sec:appendix-goal-full}

The main evaluation synthesizes goal-level commands from prior task trajectories and evaluates the augmented interface on 368 tasks across eight applications (Table~\ref{tab:method-cross-app}).

The gains land where tasks most often share multi-step subgoals: Blender improves by 5.7 points (a 12\% relative gain), Calc by 5.2, and QGIS by 3.7. The five remaining applications are unchanged because few subsequences recur across their requests, so no goal-level command was promoted for them. By design, the layer extends the interface only where prior use supplies evidence for a shared goal. On tasks that invoke a goal-level command, input tokens fall by 6.2\%. Since commands are synthesized from served requests, these figures reflect a small initial trajectory pool, and coverage of the layer grows as the interface is used.
This evaluation measures interface extension from prior use and Appendix~\ref{sec:appendix-goal-holdout} reports a separate 7:3 split for command synthesis and evaluation.

\subsection{Held-Out Evaluation}
\label{sec:appendix-goal-holdout}

To examine whether these commands can be used on tasks outside those used for synthesis, we separate the tasks used to generate commands from those used for evaluation.

\paragraph{Protocol.}
We partition the 368 tasks across eight applications into training and test sets at an approximately 7:3 ratio within each application, yielding 258 training tasks and 110 held-out test tasks in total. Using only training-task trajectories, following the goal-level command synthesis procedure described in Section~\ref{sec:method-goal}, we synthesize and retain 7 goal-level commands: 4 for Calc, 2 for QGIS, and 1 for Blender. For the remaining applications, no candidates derived from the training trajectories meet the admission criteria, so no goal-level commands are retained.

\begin{table}[htbp]
\centering
\setlength{\tabcolsep}{12pt}
\renewcommand{\arraystretch}{1.12}
\caption{Results on 110 held-out tasks.}
\begin{tabular}{lrrr}
\toprule
\textbf{Metric} & \textbf{Atomic} & \textbf{Atomic + goal-level} & \textbf{Change} \\
\midrule
Success rate & 64/110 (58.2\%) & 67/110 (60.9\%) & $+2.7$ points \\
Tokens per task & 22.8K & 21.7K & $-4.8\%$ \\
\bottomrule
\end{tabular}

\label{tab:appendix-goal-holdout}
\end{table}

\paragraph{Interpretation.}
Table~\ref{tab:appendix-goal-holdout} shows that adding goal-level commands increases successful tasks from 64 to 67 out of 110, a gain of 2.7 percentage points, while reducing tokens per task from 22.8K to 21.7K (4.8\%). This pattern is consistent with the results obtained when goal-level commands are synthesized from the full task set, demonstrating the generalizability of this approach.

\FloatBarrier

\section{Interface Generation Time}
\label{sec:appendix-generation-time}

Table~\ref{tab:appendix-generation-time} reports interface generation times for \methodname{} across eight applications, measured as the elapsed wall-clock time from when the coding agent starts generation to when it completes the process. Generation takes 17.1--36.4 hours per application.

\begin{table}[htbp]
\centering
\setlength{\tabcolsep}{4pt}
\renewcommand{\arraystretch}{1.12}
\caption{Interface generation times of \methodname{} across eight applications.}
\resizebox{\linewidth}{!}{
\begin{tabular}{lrrrrrrrr}
\toprule
& \textbf{QGIS} & \textbf{Writer} & \textbf{Blender} & \textbf{GIMP} & \textbf{Calc} & \textbf{Impress} & \textbf{Inkscape} & \textbf{FreeCAD} \\
\midrule
Generation time (hours) & 18.2 & 34.2 & 27.0 & 20.4 & 22.1 & 36.4 & 17.1 & 23.4 \\
\bottomrule
\end{tabular}
}
\label{tab:appendix-generation-time}
\end{table}

\FloatBarrier

\section{Evaluated Software Versions}
\label{sec:appendix-software-versions}

For each application, we take the newest release available when we built \sysname{} and earlier releases spaced at approximately uniform time intervals. The table below lists the exact releases evaluated for each application. We keep the surrounding runtime and evaluation environment fixed, making the installed application release the only manipulated factor.

Table~\ref{tab:appendix-software-versions} records the releases used in Section~\ref{sec:study}. The LibreOffice row applies to Writer and Impress.

\begin{table}[htbp]
\centering
\setlength{\tabcolsep}{12pt}
\renewcommand{\arraystretch}{1.15}
\caption{Software releases evaluated in the cross-version experiments.}
\begin{tabular}{ll}
\toprule
\textbf{Application} & \textbf{Evaluated releases, oldest to newest} \\
\midrule
QGIS    & 3.10.14, 3.22.16, 3.34.6, 3.44.9 \\
Blender & 2.79, 2.83.13, 3.6.23, 4.2.9 LTS \\
FreeCAD & 0.20.2, 0.21.2, 1.0.2, 1.1.1 \\
Inkscape & 1.0.2, 1.2.2, 1.3.2, 1.4.3 \\
LibreOffice & 6.4.7.2, 7.2.7.2, 7.6.7.2, 24.8.7.2 \\
\bottomrule
\end{tabular}

\label{tab:appendix-software-versions}
\end{table}

\FloatBarrier

\section{Task Counts by Application}
\label{sec:appendix-task-counts}

\begin{table}[htbp]
\centering
\setlength{\tabcolsep}{5pt}
\caption{Task counts per application in \sysname{}. Reused tasks are retained from previous benchmarks while synthesized tasks are from our pipeline.}
\renewcommand{\arraystretch}{1.1}
\begin{tabular}{lccccccr}
\toprule
\cmidrule(lr){2-3} \cmidrule(lr){4-4} \cmidrule(lr){5-5} \cmidrule(lr){6-6} \cmidrule(lr){7-7}
& \textbf{Writer} & \textbf{Impress} & \textbf{Inkscape} & \textbf{Blender} & \textbf{QGIS} & \textbf{FreeCAD} & \textbf{All} \\
\midrule
Reused       & 21 & 44 & 19 &  0 &  0 & 18 & 102 \\
Synthesized  & 35 &  4 & 12 & 35 & 54 & 13 & 153 \\
\midrule
\textbf{Total} & \textbf{56} & \textbf{48} & \textbf{31} & \textbf{35} & \textbf{54} & \textbf{31} & \textbf{255} \\
\bottomrule
\end{tabular}
\label{tab:benchmark-stats}
\end{table}

\FloatBarrier

\section{Reusing Existing Benchmarks}
\label{sec:dataset-reuse}

Our verification criterion requires every task to ship with a deterministic, artifact-based verifier. This makes OSWorld and OpenComputer natural sources of reused tasks: both provide realistic software-use tasks with executable evaluators that programmatically inspect the resulting artifact or system state rather than relying only on a model's judgment. From these benchmarks, we retain only tasks whose outcomes can be verified from the resulting artifact.

Because these benchmarks target computer-use agents, some of their tasks specify transient interface states rather than persistent artifact outcomes, and such states are not meaningfully comparable across interfaces \citep{zhou2026guivscli}. We discard two kinds: tasks that only change the application's presentation (e.g., setting the zoom to 150\% or changing an application theme) and tasks that target transient, interface-specific interaction states (e.g., restoring the slide panel in the Impress window or placing the text cursor in a Writer paragraph).

\section{Interface Synthesis Details}
\label{sec:appendix-synthesis-details}

\subsection{Atomic Command Organization}
\label{sec:appendix-command-organization}

\methodname{} organizes commands by the artifact objects they operate on and names each command by its object and action. Figure~\ref{fig:method-overview} illustrates this organization for QGIS, where commands for layouts and layers are grouped under their respective objects. An agent can browse these groups or search command names and summaries, and inspect the full specification only for commands relevant to its current task. This keeps the interface comprehensive without placing the entire command set in the agent's context.

\subsection{Goal-Level Command Construction}
\label{sec:appendix-command-packaging}

Goal-level commands package sequences that realize a shared artifact-level subgoal (Section~\ref{sec:method-goal}).
Using the atomic-command descriptions and the originating requests, a model identifies that subgoal, names the command from the shared target object and intended action following the same object-and-action convention as atomic commands, and describes its usage and resulting change. 
Details that vary across requests, such as the target object or an option value, become inputs to the new command, while the shared steps and values are held fixed. 
When invoked, the command passes these inputs to the appropriate steps and runs the steps in their original order. 
If a step fails, execution stops and the error from that atomic command is reported.

\end{document}